# Automatic Fine-grained Glomerular Lesion Recognition in Kidney Pathology


Yang Nan[a, †, *], Fengyi Li[b, †], Peng Tang[c], Guyue Zhang[b], Caihong Zeng[d], Guotong Xie[b, *],

Zhihong Liu[d,*], Guang Yang[a,*]

[a] *National Heart and Lung Institute, Imperial College London, London, UK*

[b] *PingAn Healthcare Technology, Shanghai, China*

[c] *Department of Informatics, Technical University of Munich, Munich, Germany*

[d] *National Clinical Research Center of Kidney Diseases, Nanjing University School of Medicine, Nanjing, China*


## Abstract


Recognition of glomeruli lesions is the key for diagnosis and treatment planning in kidney pathology; however, the coexisting glomerular structures such as mesangial regions exacerbate the difficulties of this task. In this paper, we introduce a scheme to recognize fine-grained glomeruli lesions from whole slide images. First, a focal instance structural similarity loss is proposed to drive the model to locate all types of glomeruli precisely. Then an Uncertainty Aided Apportionment Network is designed to carry out the fine-grained visual classification without bounding-box annotations. This double branch-shaped structure extracts



* Corresponding author: Yang Nan, Guotong Xie, Zhihong Liu, Guang Yang

Emails: y.nan20@imperial.ac.uk, xieguotong@pingan.com.cn, liuzhihong@nju.edu.cn, g.yang@imperial.ac.uk

† Denotes equal contributions






common features of the child class from the parent class and produces the uncertainty factor for reconstituting the training dataset. Results of slide-wise evaluation illustrate the effectiveness of the entire scheme, with an 8–22% improvement of the mean Average Precision compared with remarkable detection methods. The comprehensive results clearly demonstrate the effectiveness of the proposed method.

*Keywords:* Deep Convolutional Neural Network, Glomerulus Segmentation, Fine-Grained Lesion Classification, Uncertainty Assessment, Kidney Pathology



---

## 1. Introduction

Immunoglobulin A nephropathy (IgAN) is the leading cause of chronic kidney disease worldwide, especially in Asian regions, with nearly 40% of patients developing the end-stage renal disease within decades. Patients with this nephropathy have varied histological lesions, ranging from crescentic glomerulonephritis, mesangial proliferation to global and segmental sclerosis. The five main types of structure ( ) in IgAN (Fig. 1) includes Neg (tubule and arteriole), GS (Global Sclerosis), C (Crescent), SS (Segmental Sclerosis) and NoA (None of Above). However, due to the collapse and proliferation of capillary loops, some pathological changes in IgAN share a high visual similarity that even pathologists cannot achieve satisfactory agreement. A previous study found a low intra-class correlation coefficients of recognizing SS (0.66) and C (0.46) in IgAN, given by three to five pathologists [1]. Evidently, there is an urgent need to shift the balance of pathological changes' identification towards more objective and quantification.





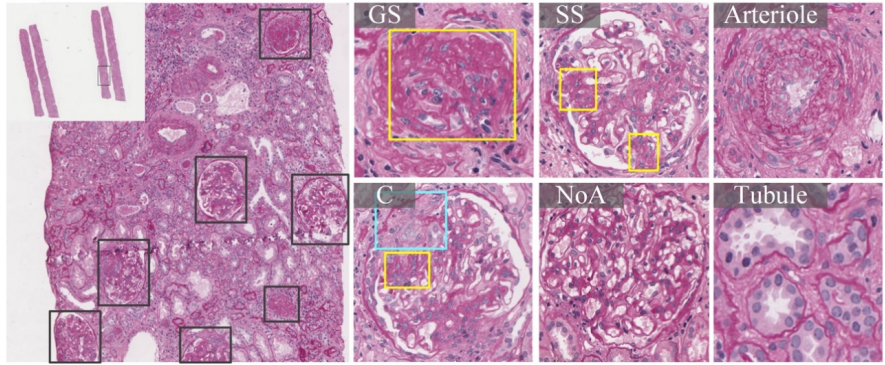

Fig. 1: Glomeruli in IgAN whole slide images with periodic acid–Schiff stain. Pathological changes of sclerosis and crescent are marked with yellow and cyan boxes, respectively.

As digital pathology evolves, biopsy tissues can be scanned as whole slide images (WSIs) through micro-scanners. Meanwhile, the remarkable success of deep convolutional neural networks also provide pathways to address the above intractable issues. With these efforts, researchers have studied computer-aided diagnosis for renal pathological feature recognition, including glomeruli location and lesion classification. Current location methods are mainly based on the combination of patch-wise predictions, including detection [2] and segmentation [3] of numerous glomeruli in large patches, or binary classification (glomerulus or non-glomerulus) in small patches [4]. However, most existing studies typically focus on classifying sclerosis and non-sclerosis and do not involve complicated lesion types, indicating the insufficient exploration of IgAN. To identify complex pathological changes in IgAN, several difficulties need to be addressed. Firstly, each glomerulus should be accurately localized. In addition, glomerular lesions with low intra-class variances should be precisely classified, even under an imbalanced data distribution. Unfortunately, most fine-grained classification tasks were performed on natural images [5, 6] with relatively unitary





background regions, which are contrary to the complex background regions in renal pathological images (Supplement-Fig. 1). Besides, methods that heavily rely on part annotations [7], such as bounding boxes or masks that label the sub-regions, are impractical for medical images due to the high cost of manual annotations. For fine-grained recognition, the network should assess the 'confidence' of its prediction to express the certainty of its output. Unfortunately, methods such as Bayesian deep learning and Monte Carlo dropout require repetitious inferences, which are inconvenient and time-consuming. Last but not least, the network applied in medical image analysis should be able to reduce the negative impact of inconclusive annotations.

In this paper, we present the first attempt to conduct fine-grained visual recognition (FGVR) in large-scaled whole slide images, aiming to recognize complex pathological changes in IgAN. Different from previous works, FGVR aims to locate but distinguish fine-grained subcategories. To address the weak capacity of existing detection modules (such as Mask R-CNN [8], FCOS [9], etc.) on FGVR task, we propose a two-stage scheme, with glomerulus segmentation and classification, respectively. Initially, a focal instance structural similarity (FISS) loss is presented to acquire accurate segmentation results. It coalesces focal loss with instance structural similarity loss to accurately segment the boundary of the glomerulus. Then, the glomerular lesions are divided into two groups based on pathological representation and lesion severity, with Neg, GS, CSN (the combination of C, SS, and NoA) as a parent class and C, SS, NoA as a child class. Based on this definition, an Uncertainty-Aided Apportionment Network (UAAN) is proposed to classify complex pathological





changes in IgAN, yielding two groups of predictions (corresponded to the parent and child classes) and their uncertainty. This uncertainty indicates the confidence coefficient of the prediction of the proposed network, which can be further applied to data reconstitution, including seeking missing annotations, mislabeled, and hard samples. With this indicator, the uncertain annotations can be picked up, rechecked, and analyzed by experts. To better illustrate the mechanism of UAAN and the interpretability, heatmaps from different layers are visualized using Gradient weighted Class Activation Mapping (Grad-CAM) [10]. Besides, uniform manifold approximation and projection (UMAP) [11] is applied to demonstrate the data distribution before and after data reconstitution. Experimental results on the Warwick-QU dataset [12] and our in-house renal pathology dataset are reported, with a patch-wise and a slide-wise evaluation respectively.

The main contributions of our work are:

- We have introduced a scheme for fine-grained visual recognition in kidney pathology, aiming to detect complex pathological changes in IgAN.

- We have proposed a focal instance structural similarity loss to improve segmentation performance by assessing the structural integrity through the instance.

- We have designed an effective architecture for fine-grained classification and uncertainty assessment, with detailed ablation experiments and heatmap visualization.

- Comprehensive experiments have been conducted on multi-levels (patch-wise and slide-wise) to prove the effectiveness of our proposed network.





## 2. Related work

### 2.1 Loss functions for accurate segmentation

The common loss functions mainly consist of Dice coefficient loss, cross-entropy loss, Jaccard loss, and focal loss [13]. In addition to the well-known losses, Tversky loss [14] was proposed as an extension to the Jaccard loss, which restricted false positive and false negative rates through hyperparameters $\alpha$ and $\beta$. Lovasz loss combined Jaccard loss with Lovasz extension to find the minima of the submodular function [15]. Structural similarity was also introduced as an optimization function for segmentation based on sliding windows [16]. However, existing studies for segmentation losses barely assess the predictions through instance level [17]. Specifically, the loss function should be designed considering each object within the image respectively, rather than adopting the same strategy for all the objects. In other words, the penalty of missing a small object should be higher than missing an equivalent area within the large object.

### 2.2 Fine-grained visual classification

The fine-grained visual classification (FGVC) was first defined to classify different species of birds from images with a single object [18, 19]. It could be divided into region-based and feature expression-based methods. The region-based approaches [20, 21] predominantly involved independent locations and fine-grained feature learning (forcing the classification model to learn discriminative features within specific regions). The feature expression-based methods normally required complex architectures or custom operations. Wu et al. [22] built a multi-path model consisting of M+1 branches (M is the number of stain modalities), aiming to



extract the complementary features from different stain modalities of a certain image. Ji et al. [23] incorporated convolutional operations along the edges of the tree structure and used the routing functions in each node to determine the root-to-leaf computational paths within the tree. However, due to the complex and various background samples, methods on natural images could barely achieve satisfactory performance in pathological studies.

## 3. Method

### 3.1 Overview

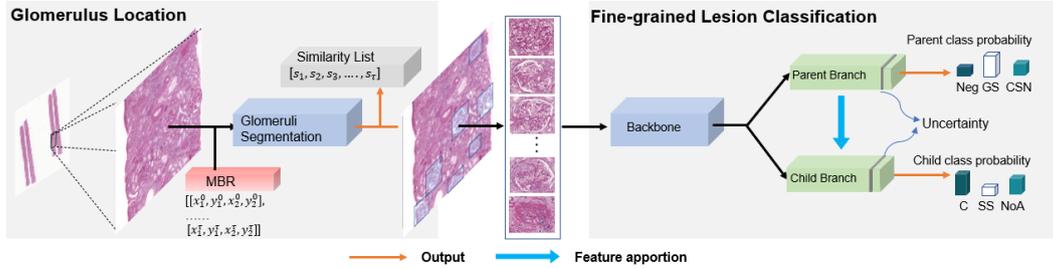

Fig. 2. Overview of the proposed scheme. The fine-grained detection task is divided into two sub-tasks, including glomerulus location and fine-grained classification.

Given a whole slide image $\mathcal{H} \in R^{W \times H \times 3}$ that can be divided into $N_\mathcal{P}$ large patches $\mathcal{P} \in R^{w \times \hbar \times 3}$ and binary mask $\mathcal{M} \in R^{w \times \hbar}$, the target bounding boxes $\mathcal{B} \in R^{n_l \times 4}$ are first extracted to train the segmentation module $\mathcal{G}_\omega$. Optimized by the proposed Focal Instance Structural Similarity loss ($\mathcal{L}_{FISS}$), $\mathcal{G}_\omega$ aims to locate all types of the glomeruli for fine-grained classification. Then the classification module $\mathcal{D}_\theta$ is trained to predict the categories $\mathcal{J}$ of all segmented objects and produce the corresponding confidence coefficient $\mathcal{C}$ as

$$\mathcal{J}, \mathcal{C} = \mathcal{D}(\mathcal{G}(\mathcal{P}, \mathcal{B}; \omega), \mathcal{F}, \ell_{CSN}; \theta) \tag{1}$$

where $\mathcal{F}$ denotes the fixed samples, $\ell_{CSN}$ indicates CSN index lists (details can be found in





3.3.1), $\omega$ and $\theta$ are the parameters of the segmentation module $\mathcal{G}_\omega$ and classification

module $\mathcal{D}_\theta$. During the inference, the glomeruli are extracted by the segmentation model,

instead of ground truth annotations.

## 3.2 Accurate segmentation of glomeruli using FISS

In this study, the segmentation module $\mathcal{G}_\omega$ is modified based on SegNet [24], introducing the

group normalization and leaky relu. Currently, most segmentation losses are designed based

on overlap measurements, while it makes the network only focus on the correct ratio of

predicted pixels to the ground truth. To address this issue, a compound loss (DL+FISS) that

considers overlap and regional structural similarity is introduced, which can be easily adopted

in various architectures. The FISS loss combines focal loss $L_F$ with instance structural

similarity (ISS) loss $L_{ISS}$

$$L_{FISS} = \alpha \cdot L_F(p,g) + \beta \cdot L_{ISS}(p,g,\mathcal{B}) \qquad (2)$$

where $\alpha$ and $\beta$ are constraint weights, $p$ and $g$ represent the prediction and ground truth, $\mathcal{B}$

indicates the bounding box of each target in the ground truth. Since we want to balance the

constraint from $L_F$ and $L_{ISS}$, we set both $\alpha$ and $\beta$ to 1, which was also proved appropriate

in our initial pilot study. The Instance Structural Similarity (ISS) loss is inspired by [25],

aiming to assess the structural integrity of each target

$$L_{ISS} = 1 - (ISS_P + ISS_N), \qquad (3)$$

where $ISS_P$ is the instance similarity of the image with Region of Interests (ROI) and $ISS_N$

indicates the similarity of background samples (images without ROI). For images with multi-

objects, ISS first evaluates the similarity between the prediction and the ground truth of each





object. Then, a list that includes similarity indexes of instances is acquired, e.g., list $[I_0, I_1, I_2, I_3, I_4, I_5]$ will be acquired when there are six instances within the input image (shown in Supplement-Fig. 2). Let $n_I$ be the number of instances in an input image $x$, $\mu_i$, $\sigma_i$ be the mean and variance of $i$-th instance within the $x$, the $ISS_P$ can be given through

$$ISS_P = \frac{1}{n_I}\sum_{i=1}^{n_I}\frac{\left(2\mu_{p_i}\mu_{g_i}+c_{1_i}\right)\left(2\sigma_{p_ig_i}+c_{2_i}\right)}{\left(\mu_{p_i}^2+\mu_{g_i}^2+c_{1_i}\right)\left(\sigma_{p_i}^2+\sigma_{g_i}^2+c_{2_i}\right)}, \tag{4}$$

where $p_i$ and $g_i$ indicate the region of the *i-th* instance extracted from the prediction and the ground truth of $x$. Instances that are smaller than the sliding window during similarity calculation are applied with zero paddings to ensure the kernel can slide across the object. Image without ROI is divided into $\lambda$ patches (we set $\lambda$ to 4) to compute $ISS_N$ instead of evaluating the similarity directly

$$ISS_N = \min[SSIM(G_1, P_1), \dots, SSIM(G_\lambda, P_\lambda)]. \tag{5}$$

By combining $L_{ISS}$ and $L_{Focal}$, it will focus on the imbalanced samples while maintaining the structural integrity, which can output smoother boundaries and fewer false negative samples. Further exploration of FISS is illustrated in Supplement-3.

### 3.3 Fine-grained lesion classification via UAAN

This section introduces the details of UAAN ($\mathcal{D}_\theta$), including hierarchical structure design, feature apportionment mechanism, uncertainty aided data reconstitution, and solutions for imbalanced dataset.

### 3.3.1 Hierarchical structure design

Deep networks can achieve superior results when categories are independent and identically distributed. However, due to the complex pathological changes in IgAN, lesions cannot be





well classified through normal solutions. In this section, we design a hierarchical architecture for fine-grained classification, with the parent branch $B_p$ for classifying classes with large variation and child branch $B_c$ for dividing class with small variation (shown in Fig. 3).

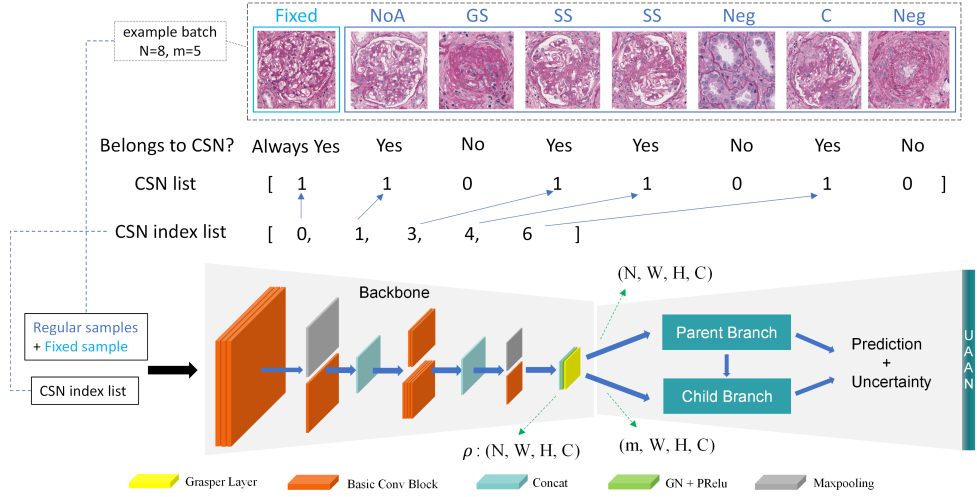

Fig. 3. The schematic diagram of the uncertainty-aided apportionment network. Each input mini-batch includes regular samples, a fixed sample, and a CSN index list.

The backbone of UAAN is inspired by the Inception-ResNetV2 [26] by introducing Group normalization, PReLU [27] (to strengthen the generalization ability), and basic convolution block (including convolution layer, GN, and PReLU). After the backbone, feature maps are separately transported to two sub-branches ($B_p$ and $B_c$) through CSN index list $\ell_{CSN}$, which is acquired during the data preparation with its elements representing the index of CSN samples in current batch. Assume a batch of N images containing m CSN samples, then $B_c \in R^{m \times w' \times h' \times c'}$, $B_P \in R^{N \times w' \times h' \times c'}$, where $w'$, $h'$, $c'$ is the number of width, height, and channel in the last feature map of the backbone. This unique mechanism requires that each batch should include at least one CSN sample.

After the backbone of UAAN, the $B_p$ and $B_c$ are performed to give parent and child class predictions, respectively. In addition, a feature apportionment mechanism is proposed between these two branches to transfer valuable information of subcategories. In both $B_p$ and $B_c$, the global maxpooling layers (GMP) are applied to extract the most intense response in each channel in the last feature map. And the dropout layers are implemented to force the network to learn more robust features that are useful in conjunction with different random subsets of other neurons. Details of the two branches can be found in Fig. 4.





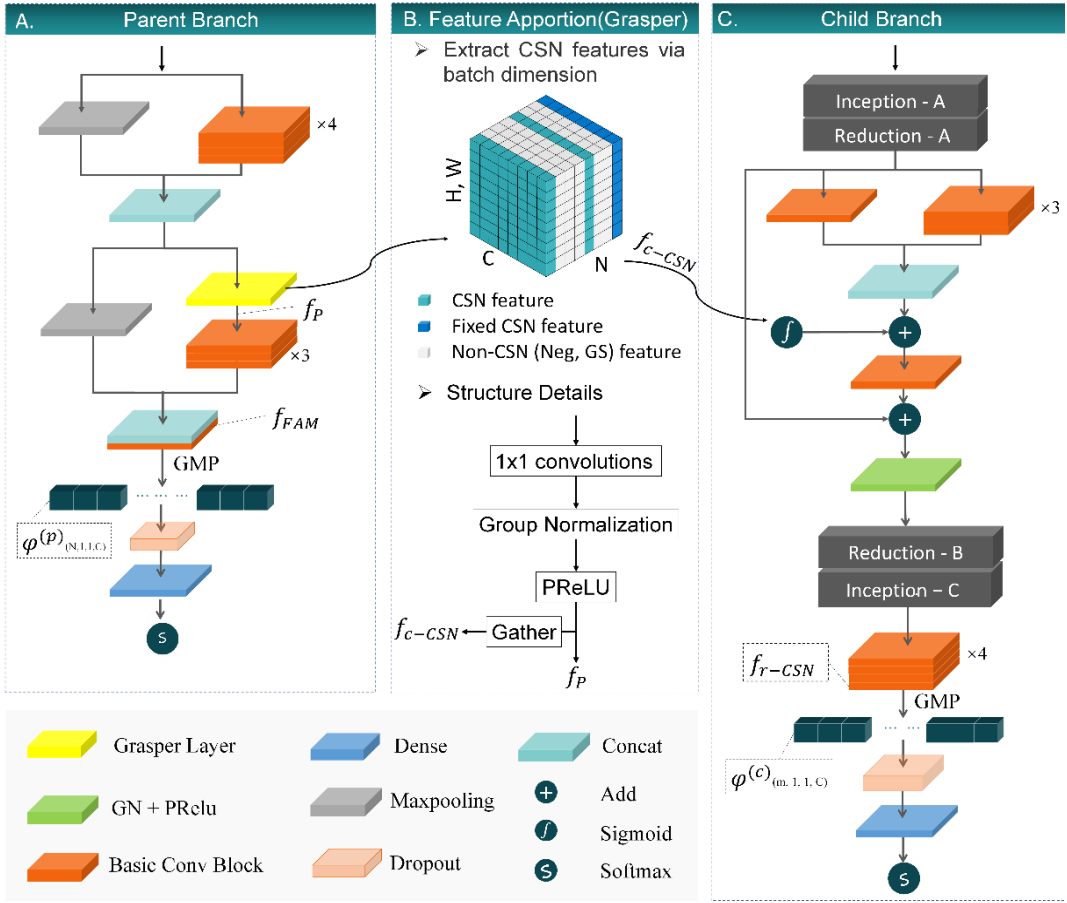

Fig. 4: The schematic diagram of the specific layers of the UAAN is shown in A. Parent branch. B. Feature apportion layer. C. Child branch.

### 3.3.2 Feature apportionment mechanism

Feature apportionment plays a crucial role in UAAN and is performed through the Grasper





layer (Fig. 4. B), aiming to extract CSN common features in the latent feature space. There are 5 classes: Neg, GS, C, SS, NoA, with a small inter-class variance between C, SS and NoA, and large inter-class variance between Neg, GS and CSN (the union of C, SS and NoA). The gather operation in the Grasper layer extracts the certain feature maps of CSN, based on $\ell_{CSN}$. Since $B_p$ is designed to classify Neg, GS, and CSN (here C, SS and NoA are regarded as a single parent class), the layers in $B_p$ aims to learn discriminative features among Neg, GS, and CSN samples. Therefore, the CSN feature in $B_p$ can be regarded as the common feature of C, SS, and NoA ($f_{c-CSN}$) that rarely exists in GS and Neg samples (e.g., capillary loops, mesangial regions). On the contrary, since $B_c$ aims to classify the C, SS, and NoA, layers in $B_c$ are optimized to extract discriminative features of child classes. Therefore, the utilization of common and discriminative features can be promoted by transferring $f_{c-CSN}$ from $B_p$ to $B_c$ through a sigmoid activation layer. This mechanism will help the network not only be trained with the discriminative features but also make use of common features.

### 3.3.3 Uncertainty aided data reconstitution

The hard samples in this study are samples that include many resembled features (or similar morphological characteristics) among C, SS, and NoA. For instance, the mesangial proliferation and crescent may co-exist in a single glomerulus, and experts usually give out their subjective judgment according to the severity degree. Due to these hard samples, even a large amount of data is provided, the network training can barely get good results. This is because including these samples in the training set makes it difficult for the network to extract features that are highly related to the real sub-category. In natural image analysis, these



*Feb. 12,2022*

'unreliable' and hard samples are usually tracked, relabeled, and augmented to enhance the capacity of computational modules. However, it is unaffordable for biomedical image analysis due to the heavy cost of expert annotation. We hold that some hard samples do not help to train a robust module, which leads to a negative swing of classification boundaries.

To bridge this gap, an uncertainty assessment is proposed to output the confidence coefficient of the prediction as to remove 'unreliable' and hard samples from the raw training set, calculated from the logits of the softmax layer $S^{(p)}(S^{(c)})$ and vectors of the GMP layer $\varphi^{(p)}(\varphi^{(c)})$ in both $B_c$ and $B_p$. Deep networks can be regarded as powerful feature extractors that extract multi-level features in different layers. The dense layer and global pooling (GP) layer are in charge of dimension reduction, while the prediction is usually given in the last dense layer by weighing up the "evidence" from the previous layer. Because the input of the softmax layer is derived from the representation of a high dimensional feature vector, we consider the output of the softmax layer (representing a probability distribution over $n$ different possible outcomes) as the estimated probability density $P_e$. Thus, the feature vectors $S^{(p)}(S^{(c)})$ are considered as the estimated probability density $P_e^{(p)}(P)_e^{(c)}$ .

$$P_e^{(p)}, (P)_e^{(c)} = \phi(\mathcal{D}(I_G, \mathcal{F}, \ell_{CSN}; \theta)), \tag{6}$$

where $I_G$ refers to each glomerulus segmented by $\mathcal{G}_\omega$, $\phi$ indicates softmax activation, $P_e^{(p)}, (P)_e^{(c)}$ represent the estimated probability density of $B_c$ and $B_p$. Besides, since the output $\varphi^{(p)}(\varphi^{(c)})$ of the last convolution layer can be regarded as the representation of $I_G$, the distribution of $\varphi^{(p)}(\varphi^{(c)})$ is close to the original distribution of the category that $I_G$ belongs to. Thus, $\varphi^{(p)}$ and $\varphi^{(c)}$ are used to calculate the raw probability density function





$P_r^{(p)}(P)_r^{(c)}$ to approximate the real probability distribution. Assuming the glomeruli feature distribution follows a Gaussian distribution, the feature vectors $\varphi^{(p)}$ and $\varphi^{(c)}$ are considered as a sampling from feature space. Therefore, the mean $\mu^{(p)}\big(\mu^{(c)}\big)$ and variance $\sigma^{(p)}\big(\sigma^{(c)}\big)$ of $\varphi^{(p)}\big(\varphi^{(c)}\big)$, and the corresponding predictions can be calculated and substituted into

$$P_r(\mu, \sigma) = \frac{\frac{1}{\sqrt{2\pi}\sigma}e^{-\frac{|k-\mu|^2}{2\sigma^2}}}{\sum_{k=0}^{K-1}\frac{1}{\sqrt{2\pi}\sigma}e^{-\frac{|k-\mu|^2}{2\sigma^2}}} \tag{7}$$

to calculate the probability density function $P$, where $K$ is the number of classes.

With the estimated probability density function $P_e$ and the raw probability density function $P_r$, the uncertainty factor can be calculated using

$$U(P_r, P_e) = 1 - \frac{\sum_{k=0}^{K-1}\big(p_r^k - \bar{P}_r\big)\big(p_e^k - \bar{P}_e\big)}{\sqrt{\sum_{k=0}^{K-1}\big(p_r^k - \bar{P}_r\big)^2}\sqrt{\sum_{k=0}^{K-1}\big(p_e^k - \bar{P}_e\big)^2}}, \tag{8}$$

where $P_r$ and $P_e$ are denoted as

$$\begin{aligned}\bar{P}_r &= \frac{1}{K}\sum_{k=0}^{K-1}p_r^k \\ \bar{P}_e &= \frac{1}{K}\sum_{k=0}^{K-1}p_e^k\end{aligned}. \tag{9}$$

With this uncertainty factor, data reconstitution can be introduced to pick out and remove the 'unreliable' and hard samples. During the training procedure, the model was first trained from scratch on the raw dataset $D_r$, and then inferenced on $D_r$ to conduct data reconstitution. The annotation is considered credible if there exists a strong correlation between $P_r$ and $P_e$, and *vice versa*. Based on the uncertainty index, we picked up 2.33% NoA, 35.91% SS, 15.15% C, and 6.42% GS in total (when the uncertainty threshold was set as 0.5). Then, those "unreliable" samples are removed from $D_r$ to reconstitute the training set





$D_c$ with the convinced samples. It is of note that the evaluation dataset remains unchanged during the entire period.

### 3.3.4 Solutions for the imbalanced dataset in classification

Details of imbalanced solution is presented in Supplement-4.

## 4. Experiments and results

### 4.1 Datasets and training strategies

The datasets in this study include Warwick-QU and our in-house Renal Pathological Whole Slide Images (RP-WSIs). The Warwick-QU dataset is used to assess the effectiveness of FISS loss (since segmentation of circular glomeruli is a relatively simple task compared to segmentation of various glands). The RP-WSIs is used to assess the effectiveness of. Details of these two sets are shown in Supplement-5.

**Warwick-QU Dataset**: The Warwick-QU [12] includes 85 training images and 80 test images, with the outer margin of glands annotated. The size of these images is $775 \times 522$ pixels.

**Renal Pathological Whole Slide Images (RP-WSIs):** 400 Periodic acid–Schiff stained WSIs of patients with biopsy-proven IgAN are selected from the National Clinical Research Center of Kidney Diseases in Jinling Hospital, ranging from 10000×30000 to 30000×50000 pixels. Essentially, 300 WSIs were used for training and validation and 100 WSIs were used for testing [28] Details of preprocessing and training strategies can be found in Supplement-6.

### 4.2 Evaluation scheme and metrics

The evaluation scheme is divided into patch-wise and slide-wise. Patch-wise experiments aim





to prove the capacity of FISS loss and UAAN, while those slide-wise are designed to demonstrate the comprehensive performance of the proposed scheme.

*4.2.1* Patch-wise evaluation

**(a) Evaluation of FISS Loss** aims to illustrate the effectiveness of the FISS loss without considering any other interferences such as network architecture, tricks, data augmentation, and unique training strategies. We implemented ablation experiments without changing the network architecture to give. The participated loss functions include Dice coefficient loss (DL), Cross-Entropy loss (CE), Focal loss (FL), Tversky loss, Lovasz loss, SSIM loss, Instance Structural Similarity loss (ISS), and Focal Instance Structural Similarity loss (FISS). Assume $P$ as the prediction given by the segmentation network and $G$ represents the ground truth, the segmentation task is assessed through Dice coefficient

$$Dice = \frac{2|P \cap G|}{|P| + |G|}. \tag{10}$$

All these losses are trained on the same network without any architectural modification.

**(b) Evaluation of UAAN** consists of five parts. Firstly, we adopt ablation experiments to analyze the effectiveness of different techniques in UAAN, including feature apportionment and uncertainty-guided data reconstitution. All these models are trained from scratch and optimized by weighted cross-entropy loss with the same parameters. Secondly, we explore the performance of different networks to illustrate the capacity of the proposed UAAN on the raw dataset $D_r$ (without data reconstitution) and convinced dataset $D_c$. Assume K as the number of classes, N as the number of whole test images. The micro-accuracy and macro-accuracy are reported as



*Feb. 12, 2022*

$$Acc_{micro} = \frac{TP + TN}{TP + TN + FP + FN}$$
$$Acc_{macro} = \frac{\sum_{i=0}^{K} \frac{TP_i}{N_i}}{K} \qquad (11)$$

where $N$ is the total number of tested images and $K$ is the number of classes, $TP_i$, $N_i$ are the number of true positive, number of samples corresponding to the i-th category, respectively. Thirdly, we visualize heatmaps of comparison models and that of the different layers in UAAN by Grad-CAM [10]. Heatmaps of the last convolution block in the parent branch $f_{FAM}$, common features of CSN $f_{c-CSN}$, and refined feature map of CSN $f_{r-CSN}$ are visualized. Fourthly, we expurgate the "unreliable" data under different uncertainty thresholds from $D_r$ and perform a second review by experts to explore the influence of different uncertainty thresholds. At last, the data distribution in latent space before and after uncertainty-aided data reconstitution is visualized by uniform manifold approximation and projection (UMAP) [11], calculated by vectors from GMP and Dense layer.

*4.2.2 Slide-wise evaluation*

The overall performance of the proposed scheme on whole slide images is assessed through the commonly used metric Average Precision (AP) and mean Average Precision (mAP). In this experiment, the ground truths (bounding boxes with classes tag) are generated through glomeruli boundaries given by experts. During the evaluation, the segmentation is first conducted to locate all glomeruli (stage-1), followed by the classification (stage-2). The inputs of classification are given by cropping the minimum bounding boxes of segmented glomeruli (given from stage-1). It is of note that all segmented objects except tiny objects (area less than 100 pixels) will be transferred to UAAN including incorrect-segmented





samples. Then, the classification module produces category predictions of these cropped rectangles, and the probability given from stage-2 is regarded as the confidence threshold in mAP computations. Objects with a certain IoU threshold are considered as true positive samples (threshold 0.5 in our experiments). We compared the $AP_{50}$ (mean average precision with 0.5 IoU thresholds) of each subcategory among our scheme and various detection methods, including Mask R-CNN [8], FCOS [9], Mask Scoring R-CNN [29], and Cascade R-CNN [30]. All these models are trained from scratch without any pre-training.

## 4.3 Results

### 4.3.1 FISS evaluation

Experimental results on Warwick-QU and RP-WSIs-P datasets for FISS loss are reported in supplement-7 through Dice coefficient score.

### 4.3.2 Uncertainty-aided apportionment network evaluation

**Ablation Study of UAAN.** The effectiveness of different techniques is presented in Table 1. It can be found that employing feature apportionment significantly improves micro-accuracy (with a 3.34% relative gain), while the uncertainty-aided data reconstitution effectively improves the macro-accuracy (with a 2.58% gain). When employing both feature apportionment and data reconstitution, there was a significant increase in both micro (5.57%) and macro (4.69%) accuracy compared with the baseline module.

Table 1. Ablation study of UAAN

| Feature Apportionment | Uncertainty | $Acc_{micro}$ | $Acc_{macro}$ |
|:---:|:---:|:---:|:---:|
| × | × | 0.8960 | 0.8677 |
| × | √ | 0.9132 | 0.8935 |
| √ | × | 0.9294 | 0.8723 |
| √ | √ | **0.9517** | **0.9146** |

**Comparison of Fine-Grained Classification.** UAAN is highly capable when dealing with fine-grained classification and outperforms all competing architectures at this task (Table 2).

Table 2. Results of fine grained glomeruli lesion classification

| Model | $Acc_{micro}$-$D_r$ | $Acc_{micro}$-$D_c$ | $Acc_{macro}$-$D_c$ |
|---|---|---|---|
| ResNet-50 [31] | 0.8495 | 0.8717 | 0.7773 |
| ResNeXt-50 [32] | 0.8294 | 0.8474 | 0.6568 |
| DenseNet-121 [33] | 0.8635 | 0.8817 | 0.7757 |
| EfficientNet-B6 [34] | 0.9092 | 0.9235 | 0.7899 |
| Inception-ResNet-v2 [26] | 0.8507 | 0.8830 | 0.7396 |
| Mean Teacher (70% annotation) [35] | / | 0.7605 | 0.6067 |
| DCL [36] | 0.8176 | 0.8373 | 0.6770 |
| SE-ResNeXt-101 [37] | 0.8943 | 0.9106 | 0.7638 |
| DenseNet+LSTM-GCNet [38] | 0.9120 | 0.9295 | 0.8263 |
| UAAN (ours) | **0.9294** | **0.9517** | **0.9146** |

The proposed UAAN achieves the best performance on both $D_r$ and $D_c$ with 95.07 % micro-accuracy (3-11% relative gain) and 91.28% macro-accuracy (12-26%), respectively. It surpasses nearly 4% than EfficientNet-B6, 5.2% than SE-ResNeXt-101, 8% than Inception-ResNet-V2 and DenseNet-121. Meanwhile, it demonstrates that by reconstituting the training





set through uncertainty factor, the performance of all comparison models has been

significantly improved, with 1% to 4% micro-accuracy gain.

**Visualization of heatmaps.** Heatmaps of UAAN and competing networks are shown in Fig.

5 using Grad-CAM. The final prediction of UAAN is marked by black bounding boxes and the

pathological changes in GS, C, and SS are annotated with yellow curves, while the "*None*" in

$f_{r-CSN}$ indicates the heat map of these two categories is not available in the child branch. It is

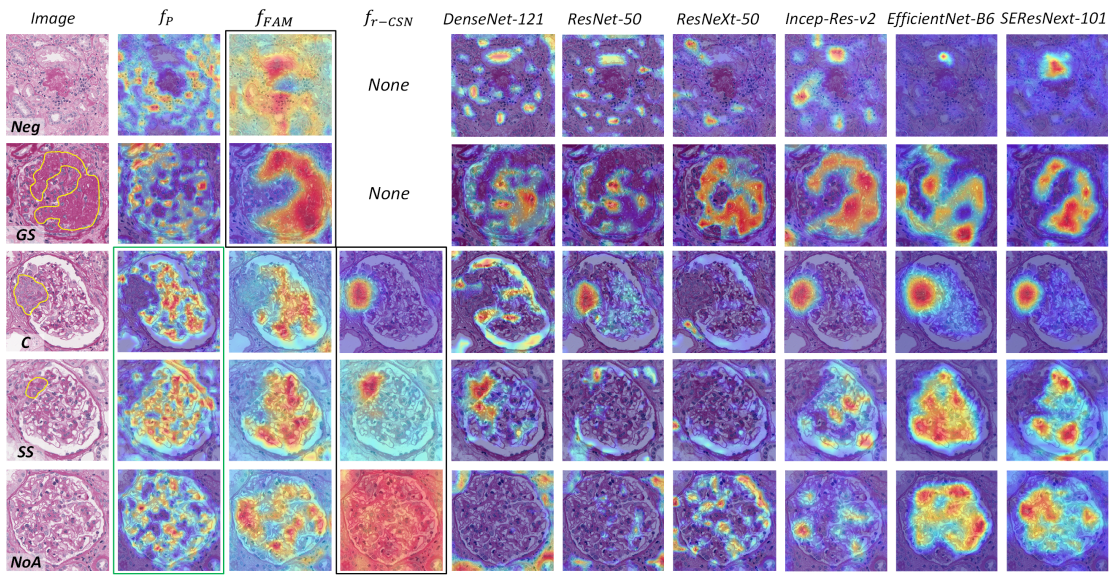

Fig. 5: Heatmap visualization of comparison models and different stages in UAAN. The heatmaps of

the final prediction of UAAN are marked by black boxes and heatmaps of $f_{c-CSN}$ are marked by the

green rectangle.

obvious that heat maps of $f_{c-CSN}$ (marked by the green rectangle) are highly related to the

CSN common features, highlighting the capillary loops, intrinsic cells, and mesangial matrix.

During the training procedure, these features are filtered by the grasper layer and conveyed to

the child branch to enhance the common information.

The highlight regions in $f_{FAM}$ straightly reflect where the network is "focusing", which





directly affects the predictions among Neg, GS, and CSN. For instance, the highlight regions in $f_{FAM-NoA}$, $f_{FAM-SS}$, and $f_{FAM-C}$ completely focus on the capillary loops, while the $f_{FAM-GS}$ focuses on the mesangial region and $f_{FAM-Neg}$ concentrates on tubules and arteriole. Relevantly, these structures are exactly the main characteristics of distinguishing Neg, GS, and CSN by pathologists. Since the output of $B_{child}$ only contains C, SS, and NoA, heat maps of Neg and GS are not available in $f_{r-CSN}$ and are marked with *None* in Fig. 5. When observing $f_{r-CSN}$, the highlight areas appear high variance, with $f_{r-CSN-C}$ focusing on the proliferating cells in the Bowman's space, $f_{r-CSN-SS}$ concentrating on the increased mesangial matrix and collapsed capillary loops, and $f_{r-CSN-NoA}$ paying attention to global information (the network needs to check whether there exist pathological changes in each corner within NoA samples).

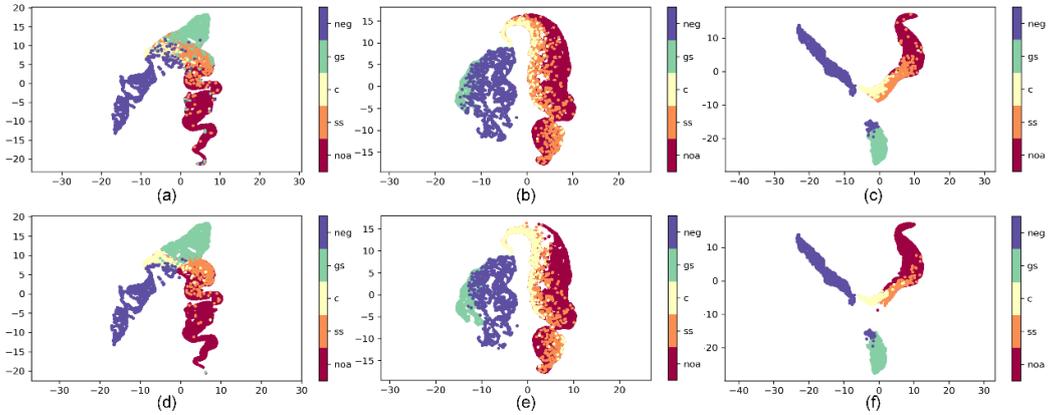

Fig. 6: Visualization of data distribution before (a-c) and after (d-f) data reconstitution via UMAP. (a) and (d) adopt the estimated probability vector from the last dense layer, (b) and (e) refer to raw probability vector from the GMP layer, (c) and (f) include both estimated and raw probability vector.

**Ablation Study of Uncertainty Thresholds.** The ablation study of different uncertainty



*Feb. 12, 2022*

thresholds is illustrated in Supplement-8.

**Effectiveness of Data Reconstitution.** Fig. 6 (a-c) shows the data distribution before

reconstitution, which is disordered, with certain NoA, SS, and C densely mixed. From

observation in Fig. 6 (e-f), employing data reconstitution via uncertainty factor tends to tear

off those mislabeled or hard samples in feature space, especially for SS, NoA, and C. While

for those low variance categories, such as GS and Neg, are less affected. The distribution map

in Fig. 6 (f) indicates that the combination of probability vectors from GMP and dense layer

exerts the clearest boundary, and the correlation between these vectors builds a bridge for

uncertainty index.

*4.3.2 Slide-wise evaluation*

Table 4 presents the results of the slide-wise evaluation for fine-grained glomeruli detection

on 100 whole slide images. The visualization results of slide-wise evaluation are presented in

Supplement-9.

Table 4. Performance of Glomeruli Detection

| Model | $AP_{NoA}$ | $AP_{SS}$ | $AP_{GS}$ | $AP_C$ | $mAP$ |
|---|---|---|---|---|---|
| Mask R-CNN [8] | 0.705 | 0.103 | 0.707 | 0.240 | 0.439 |
| FCOS [9] | **0.789** | 0.280 | 0.751 | 0.336 | 0.539 |
| MS R-CNN [29] | 0.674 | 0.110 | 0.555 | 0.230 | 0.392 |
| Cascade R-CNN [30] | 0.785 | 0.165 | **0.780** | 0.194 | 0.481 |
| Our Method (full 2-stage) | 0.780 | **0.394** | 0.718 | **0.575** | **0.613** |

Experimental results show that powerful detectors such as Mask R-CNN, FCOS, MS R-





CNN, and Cascade R-CNN present weaker performance than the proposed method (0.613 mAP), with an 8-22% reduction of mAP. To further analyze the evaluation results, the Wilcoxon-Mann-Whitney test is adopted. Statistical results show that there is no significant differences (p-value>0.05) between the proposed method and the above two when identifying GS (p-value=0.0727 for FCOS & ours, p-value=0.0619 for Cascade R-CNN & ours) and NoA (p-value=0.0895 for FCOS & ours, p-value=0.0978 for Cascade R-CNN & ours) samples. Though FCOS and Cascade R-CNN appear similar performance to the proposed method in identifying NoA and GS samples, they show poor capacity in distinguishing SS and C. By contrast, the proposed method shows balanced performance and can effectively detect all kinds of glomerular lesions, especially for SS and C.

## 5. Discussion

This section provides discussions on the proposed method, including (1) quantitative analysis of segmentation losses (2) fine-grained classification in pathological images (3) uncertainty assessment in the deep convolutional neural network, and (4) the limitation when current detection-based methods dealing with FGVR tasks.

***Segmentation Loss Functions.*** We have compared different loss functions with the FISS loss by evaluating the Dice score on simulated experiments, Warwick-QU and renal pathology datasets, respectively. Supplement-Fig.3 shows that the proposed compound loss is the most sensitive one to different sizes of targets. Supplement-Table 4 indicates the model trained with the proposed compound loss can achieve better performance among mainstream loss functions, including Dice loss, cross-entropy loss, focal loss, Tversky loss, Lovasz loss, and





SSIM loss. Although training with Dice loss and other aided losses could effectively suppress false-positive samples, it also leads to more false-negative samples, while applying DL+FISS loss can achieve an equilibrium, performing better boundaries and smooth predictions. Although there is a slight reduction of the structural integrity, employing DL+FISS actually achieves better performance compared to using the DL+ISS only.

***Fine-Grained Classification in Pathological Images.*** The ablation study in Table 1 reminds researchers to make better use of common features between different categories, instead of only forcing the network to focus on discriminative regions. As Table 1 shows, feature apportionment mainly helps the network to distinguish NoA from C and SS, while classifying C and SS still requires a cleaner data constitution. Through the reconstitution of the training set, the classification module can find a better boundary to distinguish fine-grained classes. By integrating uncertainty-aided data reconstitution with a feature apportionment mechanism, both the micro and macro accuracy are significantly improved. In Table 2, the proposed method has gained a competitive edge over other basic models, with 95.07% micro-accuracy (3–11% relative gain) and 91.28% macro-accuracy (12–26%). This huge gap has indicated the poor classification ability of current competitive models [26, 31-34, 36, 37] when dealing with data imbalanced issue. Compared with most fine-grained classification methods, the proposed UAAN has several advantages: (1) It has not required a large number of branches or generative models (2) It has considered both discriminative features and common features between subcategories, instead of focusing on the discriminative part only (3) It has a streamlined structure and can be further improved by adding complex operations such as





atrous spatial pyramid pooling and attention module. (4) It has not required any part annotations or subnetworks.

***Interpretability of UAAN.*** The visualization of heat maps at different layers has proved the reasonability of our architectural design. Besides, we have presented heatmaps of the comparison models in terms of different glomerular lesions. Among these five categories, SS has been found as the most difficult class for recognition, with only UAAN and DenseNet-121 focusing on the right place. As Fig. 5 illustrated, even some state-of-the-art models exert high accuracies (such as SE-ResNeXt-101, EfficientNet-B6), their attention regions have not been fully concentrated on the pathological changes. In contrast, UAAN's attention is much closer to pathologists, performing high interpretability of the prediction.

***Uncertainty in Deep Convolutional Neural Network.*** The uncertainty is quite important in deep convolutional neural networks (especially in medical applications) since it reflects the confidence coefficient of the current predictions. In our study, we have introduced the uncertainty assessment by calculating the correlation between two probability densities. Although our assessment has not provided a strict uncertainty calculation using Bayesian inference, it can still estimate the confidence coefficient of the predictions to some extent. Besides, it has also been used for selecting low-annotation data and hard samples, which has not required multiple forward propagations. The ablation experiments in Table 3 show that, with the increase of uncertainty threshold, fewer samples have been selected while the ratio of the mislabeled samples is higher. With detailed inspection by three senior pathologists, 36% and 55% among the mislabeled images have been corrected as C and SS (in case of



*Feb. 12,2022*

$U_{thresh} = 0.5$). This has proved our assumption that the output of the GMP layer is more correlated to the real distribution and can be used to simulate the raw probability density function. If there is no obvious correlation between $P_r$ and $P_e$, we consider the features are misrepresented in the last dense layer. That is, the output of the softmax layer represents an erroneous probability distribution due to the inaccurate feature representation process. It might be the reason why CNN+SVM is better than pure CNN in some previous studies [39]. Aided by this uncertainty assessment, significant improvements (1–4%) of micro-accuracy have been achieved by current state-of-the-art models [26, 31-34, 36, 37] using data reconstitution.

***Comparisons with junior pathologists.*** We compared the performance of UAAN with a junior pathologist on a pilot dataset (200 images per class). The labels of this pilot dataset were given by two senior pathologists, with the inter-observer agreement of Cohen's Kappa = 0.791, p < 0.0001. The junior pathologist achieved 89.4% accuracy compared to the 90.7% of UAAN's, while the UAAN (7.57ms per glomerulus on Nvidia Tesla V100 16GB, batch size 32) is nearly 260 times more efficient compared to a junior pathologist (2s per glomerulus), which shows the feasibility of conducting pathological diagnosis by neural networks.

***Why two stages?*** Though the current state-of-the-art detection modules can achieve competitive performance for locating all types of glomeruli compared with the two-stage scheme, these modules cannot well classify the fine-grained lesions. One main reason is the resolution. Due to the scale of viewings, the detection module can only be well performed under 4X resolution (2.468 μm/pixel) with an input size of 2048*2048, however, under





certain resolution the model cannot extract fine-grained features to distinguish fine-grained subcategories. Moreover, adding complex classification heads accounts for a huge module that leads to the shortage of GPU memory.

***Limitations.*** Although our work has achieved superior performance among the current methods, it has some limitations. The proposed two-stage scheme suffers from a complex feature extraction procedure, and the performance of stage-2 (fine-grained classification) is highly related to that of the segmentation stage. For instance, the classification module cannot work well when two contiguous glomeruli cannot be segmented individually. In addition, the various annotation qualities, expert subjectivity, and dataset may affect the stability of data reconstitution. Indeed, external validation can be helpful for testing the generalizability of our proposed deep learning model, but at the moment we have no access to additional data resources. In lieu of constructing an external validation dataset, we will publish our implementation on opensource platform for reproducibility studies and external validation by other researcher groups.

## 6. Conclusion

In this study, we have introduced a comprehensive scheme for glomeruli lesion recognition, employing focal instance structural similarity loss and the uncertainty-guided apportionment network. This work is the first attempt to tackle the fine-grained visual recognition task in pathological image analysis and has achieved superior performance in IgAN whole slide images. Both the focal instance structural similarity loss and uncertainty-aided apportionment networks are effective, resulting in more than 8–22% improvement of the mAP compared





with current schemes. The proposed method provides a high-precision computational scheme for fine-grained lesion identification of IgA nephropathy in whole slide images, which helps pathologists make more objective and effective clinical diagnoses. For future work, it is of note that the detected lesions are not specific for IgA nephropathy, and one of our future research directions will be the transfer learning for the lesion's detection to other nephropathies with a similar presentation, e.g., lupus nephritis and diabetic nephropathy.

## 7. Acknowledgements


This research was funded in part by the National Key Research and Development Program of China (2016YFC0901202), the National Natural Science Foundation of China (82070793), and the Project of Invigorating Health Care through Science, Technology and Education of Jiangsu Province Medical Key Talent (ZDRCA2016098), in part by the European Research Council Innovative Medicines Initiative on Development of Therapeutics and Diagnostics Combatting Coronavirus Infections Award "DRAGON: rapiD and secuRe AI imaging based diaGnosis, stratification, fOllow-up, and preparedness for coronavirus paNdemics" [H2020-JTI-IMI2 101005122], in part by the AI for Health Imaging Award "CHAIMELEON: Accelerating the Lab to Market Transition of AI Tools for Cancer Management" [H2020-SC1-FA-DTS-2019-1 952172], in part by the MRC (MC/PC/21013), and in part by the UKRI Future Leaders Fellowship (MR/V023799/1).

*Feb. 12, 2022*

*Feb. 12, 2022*

*Feb. 12,2022*

Feb. 12,2022

*Feb. 12,2022*

*Feb. 12, 2022*

## Supplementary Materials

*1. Figures*

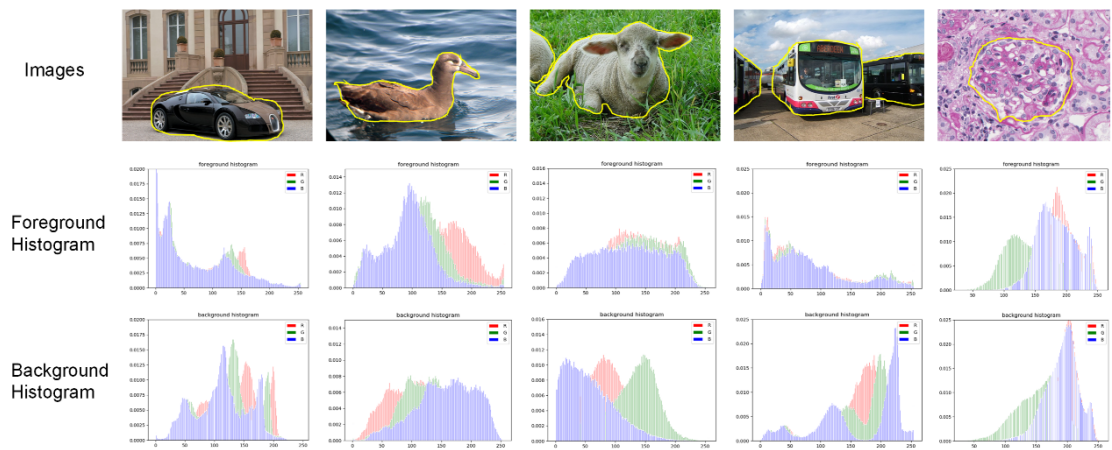

**Supplement-Fig. 1**. Histogram of images from Stanford Cars [1], CUB-200-2011 [2], PASCAL-VOC2012 [3], and our in-house dataset, with red, green, and blue region representing the intensity value from R, G, B channels respectively. The foreground objects are annotated by yellow curves.

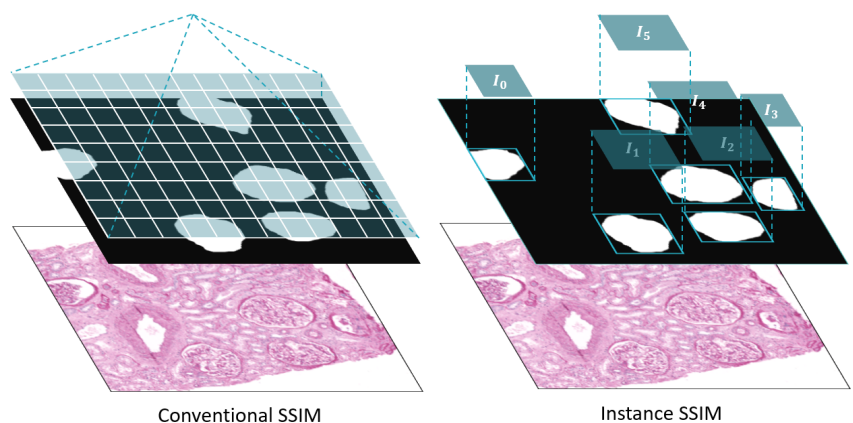

**Supplement-Fig. 2:** Computation mechanism of previous SSIM methods and the proposed ISS (instance SSIM)





*2. Definition of Five types of structures in IgAN*

Neg: Negative structures such as tubule and arteriole

Global Sclerosis: the entire glomerular tuft involved with sclerosis

Crescent: the presence of at least two layers of cells that are filling circumferential or circumscribed the Bowman's space

Segmental Sclerosis: any amount of the tuft involved with sclerosis, but not involving the whole tuft



## 3. Further exploration of FISS

To better explore the performance of segmenting a circular object through different losses, we simulate thousands of situations under different ground truth ratios (shown in Supplement-Fig. 3). Various circumstances under different ground truth ratios (reflecting the different sizes of target objects) are simulated to explore the performance of different loss functions.

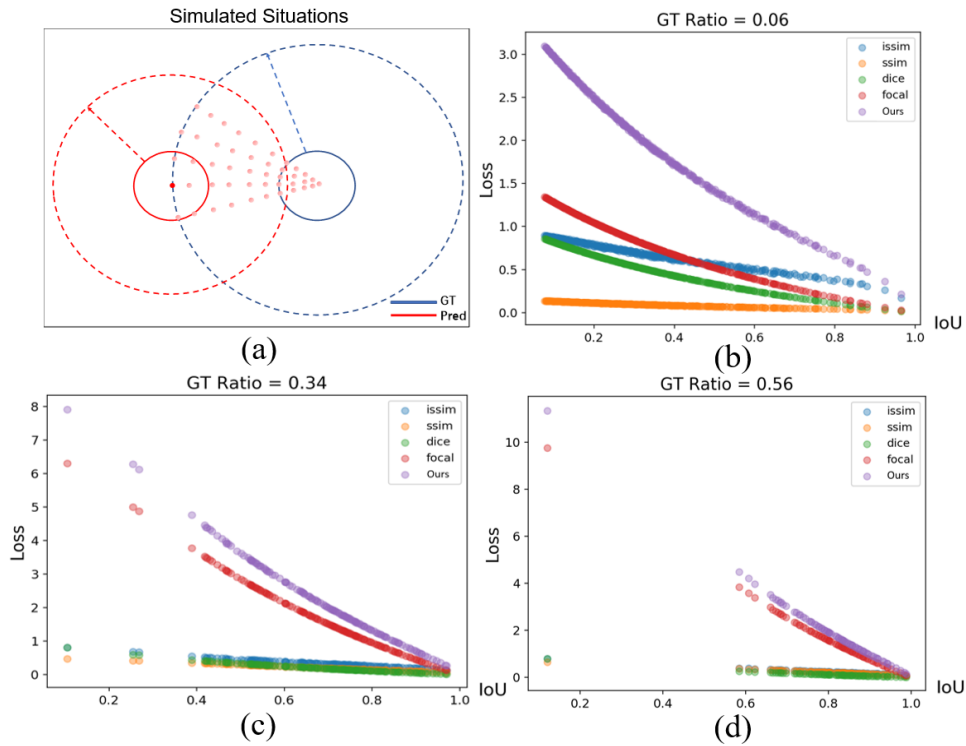

**Supplement-Fig. 3**. Analysis of different loss functions. (a) indicates the simulated situations of glomerulus segmentation, the red and blue circles can be treated as the prediction and ground truth, respectively, and the red points represent potential locations of the predicted glomerulus. (b-d) demonstrate the value of different loss functions under different segmentation quality and object sizes. 'Ours' indicates our proposed compound loss function which is the sum of dice loss, issim loss and focal loss.



Due to the updated mechanism of the network's parameters (update on batch-sized images after each iteration), an image including both small and large objects can be treated as mini-batch sub-images with a single object. Therefore, if a loss function can get a good response of a single object under different circumstances, then it must be effective for images containing various targets. Each red and blue circle can be treated as the prediction and the ground truth, respectively, and the red points can be regarded as the center of the predicted glomerulus. Initially, we introduce the ground truth ratio, reflecting different sizes of the target objects

$$GTR = \frac{a_{obj}}{a_{img}}, \qquad (S1)$$

followed by dividing images with different GTR (ground truth ratios) into three subcategories: small-sized objects (GTR not more than 20%), middle-sized objects (GTR greater than 20% and less than 50%), and large-sized objects (GTR not less than 50%). For small objects with a 6% area ratio in Supplementary-Fig. 2(b), $L_{SSIM}$ is more insensitive compared with general $L_{Dice}$ and $L_{ISS}$, which makes it less effective in small object segmentation. The $L_{ISS}$ presents a more sensitive response to small objects, even when the prediction is close to the ground truth (e.g., we empirically consider the object is well segmented if its IoU is greater than 0.8). For middle-sized objects in Supplementary-Fig. 2(c), $L_{Dice}$ and $L_{SSIM}$ present similar penalties, while $L_{ISS}$ is much stricter. For large-sized objects in Supplementary-Fig. 2(d), $L_{SSIM}$ and $L_{ISS}$ show similar effects and can precisely reflect structural integrity.



## 4. Solutions for imbalanced dataset in classification

The hierarchical structure of UAAN requires that each batch should include at least one CSN sample, which might be inapplicable for an imbalanced dataset. Therefore, we selected 3 "fixed" samples (each one from C, SS, and NoA), and randomly fed one of them into the network with regular training samples (suppose the number of input images per batch is 8, then the size of regular samples is 7). For instance, if a mini-batch contains [fixed sample, NoA, GS, SS, SS, Neg, C, Neg], then the N and m will be 8 and 5, respectively, and the 'CSN index list' will be [0, 1, 3, 4, 6] (Fig. 3). It is noted that the "fixed" samples are not involved in loss calculation and backpropagation to prevent overfitting.

In addition to the "fixed samples", we adopt a weighted cross-entropy loss for optimization, the weight index $\delta_k$ of $k$th class is calculated by

$$\delta_k = \frac{N_T / n_k}{\sum_{k=1}^{K}(N_T / n_k)} \, ,\tag{10}$$

where $K$ is the number of classes, $n_k$ is the number of the $k$th class and $N_T$ represents the total number of all classes. Therefore, the loss function for training UAAN is presented as

$$\mathcal{L} = -\sum_{k=1}^{K} \delta_k t_k log p_k \, ,\tag{11}$$

where $t_k$ and $p_k$ indicate the k-th value of the one-hot encoded label and softmax probability, respectively. Meanwhile, data augmentation is adopted due to severe class-imbalance issues, including randomized hue adjustment (randomly adjusting the hue of input images by $\Delta \in [-0.15, 0.15]$), randomized horizontal flip, randomized vertical flip, and randomized brightness adjustment ($\Delta \in (0, 0.15)$), where $\Delta$ is the scalar added to the



normalized image representation.



## 5. Details of dataset

Supplement-Table 1. Acquisition parameters of two datasets

| Dataset | Warwick-QU | RP-WSIs |
| --- | --- | --- |
| Type | Colorectal cancer | IgA nephropathy |
| Resolution | 20X (0.62005 $\mu m$/pixel) | 40X (0.2468 $\mu m$/pixel) |
| Scanner | Zeiss MIRAX MIDI | Leica Aperio ScanScope AT Turbo |
| Number of samples | 165 images | 400 whole slide images |
| Staining | Hematoxylin and Eosin | Periodic acid–Schiff |
| Number of patients | n/a | 400 |

**Warwick-QU Dataset**: There are two subsets (divided by the organizers) for testing in Warwick-QU, part A (60 images) for the off-site test and part B (20 images) for the on-site test – to some extent introduces bias into the performance evaluation of segmentation algorithms due to equal weight given to performance.

**Renal Pathological Whole Slide Images (RP-WSIs)**: 400 Periodic acid–Schiff stained WSIs of patients with biopsy-proven IgAN are selected from the National Clinical Research Center of Kidney Diseases in Jinling Hospital (collected from 2016-2019 by different operators and scanners). All data were desensitized followed by the tenets of the Declaration of Helsinki [4]. A weakly supervised self-training strategy was applied to annotate the glomeruli outer margins. The outer margins of glomeruli within 20 WSIs were first annotated by experienced



junior pathologists through Aperio Image Scope software (1.2.3.3). Then a simple UNet is trained to segment the glomeruli boundaries of all cases, followed by the revision of junior pathologists. The categories of glomeruli lesions were manually given by experienced junior pathologists, then carefully checked by three senior pathologists. The Inter-observer agreement evaluation for glomeruli lesions is (Kendall's W = 0.802, Fleiss' kappa = 0.761, p < 0.0001). With these annotations, RP-WSIs are split into three sub-datasets in terms of different scales, shown in Supplement-Table 2.

Supplement-Table 2. Composition of Renal Pathological Dataset

| Level | Training Samples | Test Samples |
|---|---|---|
| Slide (***RP-WSIs-S***) | 300 | 100 |
| Patches (***RP-WSIs-P***) | 9700 | 3700 |
| Glomeruli (***RP-WSIs-G***) | 23113 | 8529 |

\* ***P*** indicates the large patches used for the segmentation task, ***G*** represents the glomeruli patches used for the classification task, and ***S*** is the whole slide images used for the detection task.

Both the training set and test set for patches ***RP-WSIs-P*** and glomeruli ***RP-WSIs-G*** were cropped from the corresponding training and test slides, respectively. *P* included more than 13,000 large patches with the size of 1024 × 1024 pixels, while ***RP-WSIs-G*** was built through cropping the minimum bounding rectangle of each glomerulus. The distribution of glomeruli images is shown in Supplement-Table 3.

Supplement-Table 3. Glomeruli Distribution of RP-WSIs-G

| | NoA | SS | C | GS | Neg | Total |
|---|---|---|---|---|---|---|



| | | | | | |
|---|---|---|---|---|---|
| Training ($D_r$) | 10401 | 1066 | 818 | 2223 | 8605 | 23113 |
| Training ($D_c$) | 10161 | 683 | 694 | 2080 | 8409 | 22027 |
| Test | 3764 | 639 | 444 | 1017 | 2665 | 8529 |

$D_r$ refers to the raw training set and $D_c$ indicates to the 'convinced' dataset reconstituted by uncertainty factor.



## 6. Preprocessing and Training Strategies

For each input image $x$, assume $\mu$ and $\sigma$ as the mean and variance of $x$, the standard min-max normalization was employed for pre-processing through

$$x^* = \frac{\frac{x - \mu}{\sigma} - x_{min}}{x_{max} - x_{min}}. \tag{S2}$$

All trainable parameters in the experiments were initialized using the Glorot uniform initializer. Networks that participated in the comparison were trained with the Adam optimizer on an NVIDIA Tesla V100 GPU for 100 epochs, with the initial learning rate of 0.0001 and the decay of 0.96 per epoch. To present an impartial evaluation, all models were trained from scratch and did not incorporate any transfer learning strategy.



*7. FISS Loss Evaluation*

Experimental results of FISS loss are reported in S-Table 1. Performance on Warwick-QU and

RP-WSIs-P datasets are reported through the Dice coefficient.

Supplement-Table 4. Comparison Experiments for Gland Segmentation

| Loss Function | Warwick-QU | | | RP-WSIs-P |
|---|---|---|---|---|
| | $Dice_A$ | $Dice_B$ | $Dice_{A\&B}$ | $Dice$ |
| DL | 0.8512 | 0.8819 | 0.8665 | 0.742 |
| DL+CE | 0.8888 | 0.8917 | 0.8902 | 0.879 |
| DL+FL | 0.8949 | 0.8938 | 0.8943 | 0.848 |
| DL+Tversky | 0.8891 | 0.8701 | 0.8796 | 0.882 |
| DL+Lovasz | 0.8965 | 0.8683 | 0.8824 | 0.884 |
| DL+SSIM | 0.8825 | 0.8838 | 0.8831 | 0.885 |
| DL+ISS | 0.9036 | 0.8835 | 0.8935 | 0.897 |
| **DL+FISS** | **0.9049** | **0.8940** | **0.8995** | **0.906** |

S-Table 1. shows that the model trained with Dice + FISS loss achieves the best performance

(Dice: 0.9022 and 0.906) on both Warwick-QU and RP-WSIs dataset, followed by Dice + ISS

loss. We notice that there exists a large bias (the 3rd place on the Warwick-QU dataset and the

7th place on the RP-WSIs dataset) of the performance for DL&FL since it requires manual

adjustments of $\alpha$ and $\gamma$. The network trained with Dice loss performs the worst result on

these two datasets, and this might be caused by hard examples (glands with a variety of



histologic grades, arterioles like global sclerosis glomeruli). The visualization results of the

Warwick-QU dataset are reported in Supplement-Fig. 4.

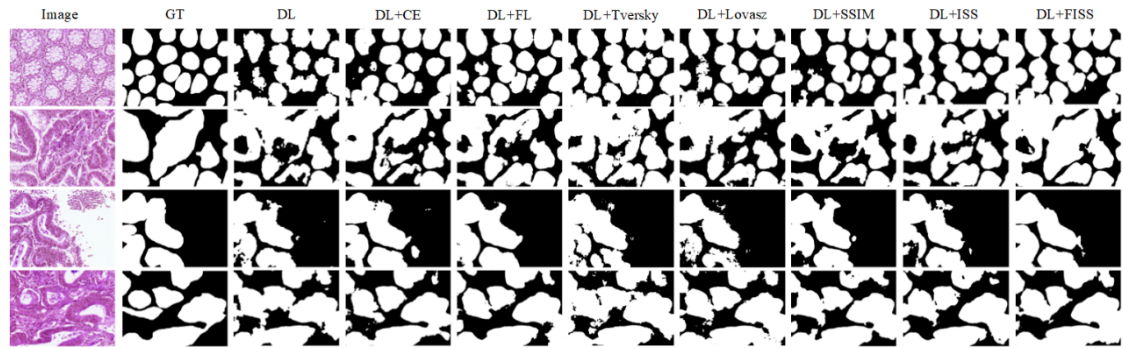

**Supplement-Fig. 4**: Visualization results of different loss functions trained on the same network from

scratch, without any preprocessing, post-processing, and data augmentation.



## 8. Ablation study of uncertainty thresholds

The initial mislabeled rate in the raw dataset $D_r$ is 6.7%, given by three senior pathologists randomly checking 100 images per class. ('/' represents no uncertainty threshold applied).

Supplement-Table 5. Ablation Experiments of Uncertainty Factor

| $U_{thresh}$ | / | 0.3 | 0.4 | 0.5 | 0.6 | 0.7 | 0.8 | 0.9 |
|---|---|---|---|---|---|---|---|---|
| $N_{select}$ | 400 | 3217 | 1729 | 1086 | 718 | 469 | 262 | 140 |
| $N_m$ | 27 | 453 | 318 | 215 | 145 | 99 | 51 | 26 |
| $R_m$(%) | 6.7 | 14.1 | 18.4 | 19.8 | 20.2 | 21.1 | 19.5 | 19.2 |

\* $N_{select}$ indicates the number of selected data from the training set $\boldsymbol{P}$ under different uncertainty threshold, $N_m$ and $R_m$ are the number and percentage of the selected data that are mislabeled, respectively. '/' represents no uncertainty threshold applied.

With the aid of uncertainty indicator $U$, 1-15% ($U \in [0.3, 0.9]$) of the training set are suspected as "unreliable" data. We expurgate these data from $D_r$ and perform a second review from experts to explore whether these images belong to hard or mislabeled samples.





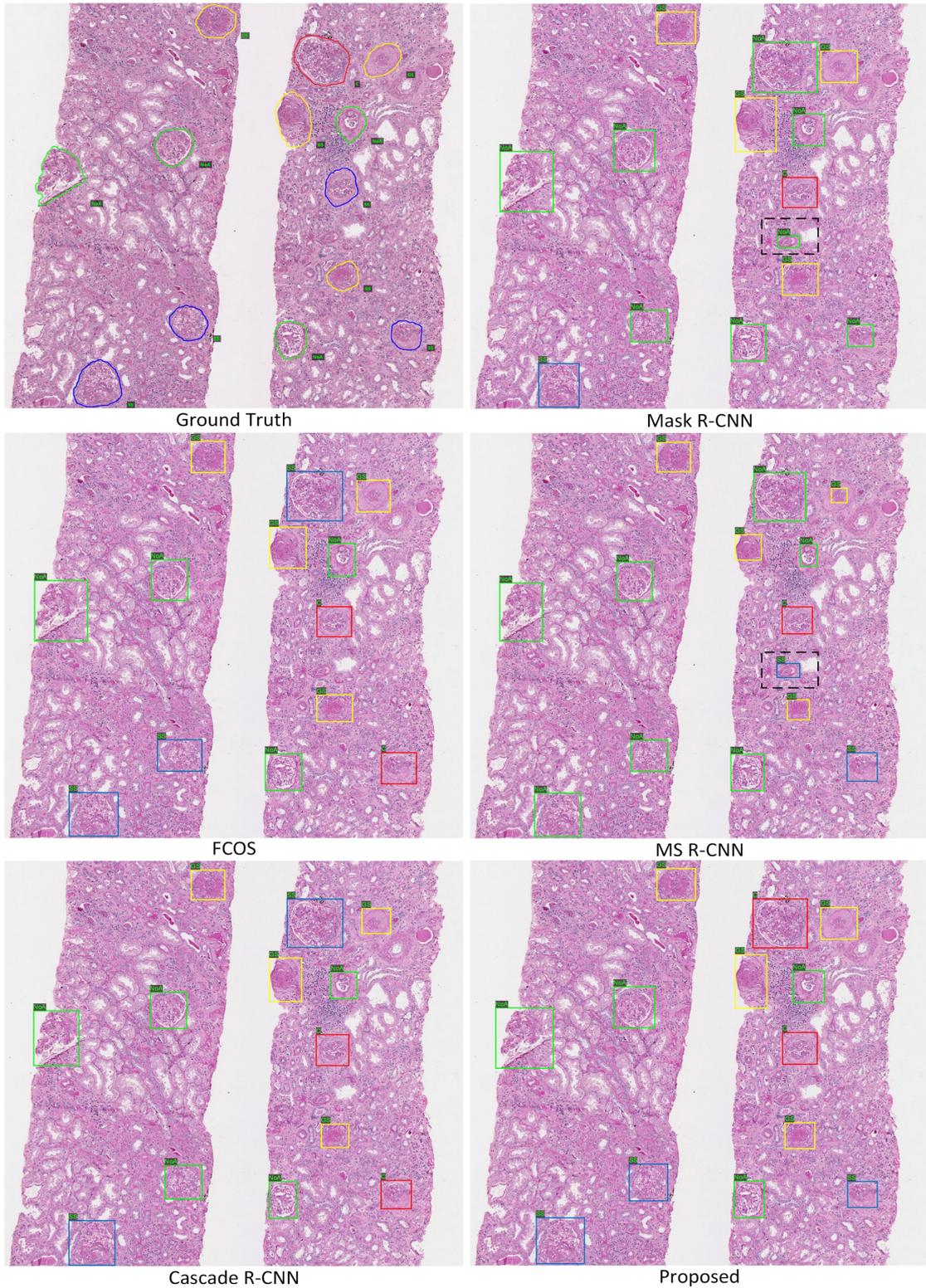

Supplement-Fig. 5 . Visualization results of the slide evaluation, the red, green, yellow and

blue boxes represent annotations/predictions of C, NoA, GS and SS, respectively. All the



arrows indicate incorrect predictions, including the black arrows (for incorrect classes), the blue arrows (for the hard sample that all models are failed to recognize) and the red arrows (indicating the incorrect foreground prediction, in which a vascular structure is recognized as a glomerulus). In addition to the Mask R-CNN and MS R-CNN who take a vascular structure as a glomerulus, all the incorrect predictions occur on SS and C, while the proposed method achieves the least incorrect predictions among all the comparison models.

Visualization results of Mask R-CNN, MS R-CNN, FCOS, Cascade R-CNN and our method are shown in Supplement-Fig. 5. All the comparison methods can clearly recognize GS, while there exert significant visual differences for detecting SS and C. Both Mask R-CNN and MS R-CNN take a vascular structure as a glomerulus (highlighted through red arrows), indicating a relatively weak performance of glomeruli detection. In addition to the Mask R-CNN and MS R-CNN who take a vascular structure as a glomerulus, all the incorrect predictions occur on SS and C, while the proposed method achieves the least incorrect predictions among all the comparison models. For instance, all methods except ours failed to recognize the unique C from the image and were struggling when identifying SS and C samples. Overall, our method achieves the best performance of detecting complex kidney lesions of IgAN.